\documentclass[runningheads]{llncs}
\usepackage{graphicx}
\usepackage{multirow}
\usepackage{multicol}
\usepackage{subcaption}
\usepackage{enumitem}
\usepackage{xcolor}
\usepackage{arydshln}
\usepackage{mathtools}

\def\barr{\begin{tabular}{l}}
\def\earr{\end{tabular}}
\begin{document}
\title{Automated Detection of Congenital Heart Disease in Fetal Ultrasound Screening}

\titlerunning{CHD Detection in Fetal Ultrasound Screening}

\author{
Jeremy Tan\textsuperscript{1*} \and
Anselm Au\textsuperscript{1*} \and
Qingjie Meng\textsuperscript{1} \and
Sandy FinesilverSmith\textsuperscript{2} \and
John Simpson\textsuperscript{2} \and
Daniel Rueckert\textsuperscript{1} \and
Reza Razavi\textsuperscript{2} \and
Thomas Day\textsuperscript{2} \and
David Lloyd\textsuperscript{2} \and
Bernhard Kainz\textsuperscript{1}}

\institute{Imperial College London, SW7 2AZ, London, UK \\
\and
 King's College London, St Thomas' Hospital, SE1 7EH, London, UK \\
\email{j.tan17, anselm.au @imperial.ac.uk}\\
}

\authorrunning{Tan J. et al.}

%
%
%
%
%
\maketitle              
\begin{abstract}
Prenatal screening with ultrasound can lower neonatal mortality significantly for selected cardiac abnormalities. However, the need for human expertise, coupled with the high volume of screening cases, limits the practically achievable detection rates. In this paper we discuss the potential for deep learning techniques to aid in the detection of congenital heart disease (CHD) in fetal ultrasound. We propose a pipeline for automated data curation and classification. During both training and inference, we exploit an auxiliary view classification task to bias features toward relevant cardiac structures. This bias helps to improve in F1-scores from 0.72 and 0.77 to 0.87 and 0.85 for healthy and CHD classes respectively.  

\end{abstract}%

\section{Introduction}
Ultrasound is the foremost modality for fetal screening. Its portability, low cost, and fast imaging make it one of the easiest imaging modalities to deploy. This gives front-line sonographers the tools to perform screening at a \textit{population} level. In each fetal examination, sonographers must inspect a wide range of anatomical features including but not limited to the spine, brain, and heart. The breadth of this task makes it difficult for sonographers to develop specialized expertise for every anatomical feature. Unfortunately, this can lead to some conditions going undiagnosed. The most fatal of which is congenital heart disease (CHD) which is associated with over 47\% of perinatal deaths and over 35\% of infant deaths (only considering deaths related to congenital abnormalities)~\cite{NCARDRS2017}. 

Experts can detect CHD's with over 98\% sensitivity and near 90\% specificity~\cite{bennasar2010accuracy,yeo2018fetal}. However, shortage of specialists means that over 96\% of examinations are performed by generalist sonographers~\cite{van2016prenatal}. As a result, \textit{population}-based studies consistently report detection rates around 39\%~\cite{pinto2012barriers} (with one exception reaching 59\%~\cite{van2016prenatal}). Machine learning methods could help close this gap and provide sonographers with assistance for more difficult diagnoses.

Deep learning has been used in many fetal ultrasound applications including standard plane detection~\cite{baumgartner2017sononet,cai2018multi,chen2015standard,kong2018automatic}, extrapolation of 3D structure from 2D images~\cite{cerrolaza20183d}, and biometric measurements for developmental assessment~\cite{sinclair2018human,kim2018machine}. However, there have been relatively few works on diagnostic assistance in fetal screening. Some of the major challenges in this application are i) data curation and ii) disease variance. Data curation is crucial because only certain ``standard planes'' are considered diagnostic~\cite{Fasp2018}. Extracting relevant frames for CHD is particularly challenging because pathological cases \textit{by definition} deviate from the description of these standard planes. The high variation of the manifestation of CHD's also make diagnosis difficult. 

In this work we propose a pipeline to perform automated diagnosis of hypoplastic left heart syndrome (HLHS), a term which encompasses a spectrum of malformations in the left ventricle and its outflow tract~\cite{simpson2000hypoplastic}. A standard plane detector is first used to extract relevant frames from the healthy and pathological cases. This data is then used to train a classifier to distinguish between normal control (NC) cases and HLHS patients. 

\section{Related Work}
Most successful applications of deep learning in fetal ultrasound have been in pre-diagnostic tasks. In particular, standard plane detection has been studied extensively\cite{baumgartner2017sononet,cai2018multi,chen2015standard,kong2018automatic}. Standard plane detection is also involved in the proposed pipeline and is based on the SonoNet architecture~\cite{baumgartner2017sononet}. 

Most closely related to our work is a recent study on diagnosing HLHS and Tetralogy of Fallot (TOF) in fetal ultrasound~\cite{arnaout2018deep}. In their methodology, images are labelled based on standard planes allowing for extraction of relevant images with high diagnostic quality. Using these images, they train a series of binary classifiers to distinguish between healthy and pathological cases. Each classifier is trained using images from only one standard plane. The predictions from each plane are then summed and a threshold is used to determine the final diagnosis. 

They achieve high sensitivity and specificity demonstrating that neural networks can learn to identify CHD. In their study, the training images are of high diagnostic quality and come from a diverse dataset including about 600 patients (for normal vs. HLHS). These favorable conditions are not always possible because of a lack of expert annotations and the rarity of CHD conditions. As such, we aim to investigate the feasibility of CHD detection in the low data regime, using a total of 100 patients. We also study the impact of automated and manual data curation. 

The design of \cite{arnaout2018deep} is also dependent on data curation. Training individual networks for each standard plane means that training data must be accurately sorted into the correct planes. It also means that test data must be accurately sorted for inference because each network is only trained to recognize pathology in a single view. Furthermore, using individual networks for each plane leads to larger memory requirements. Instead, we train a single network for all views and explore the use of plane labels as an auxiliary task for multitask learning~\cite{caruana1997multitask}. 

%

\section{Method}
\begin{figure*}[h]
	\centering
        \includegraphics[width=\linewidth]{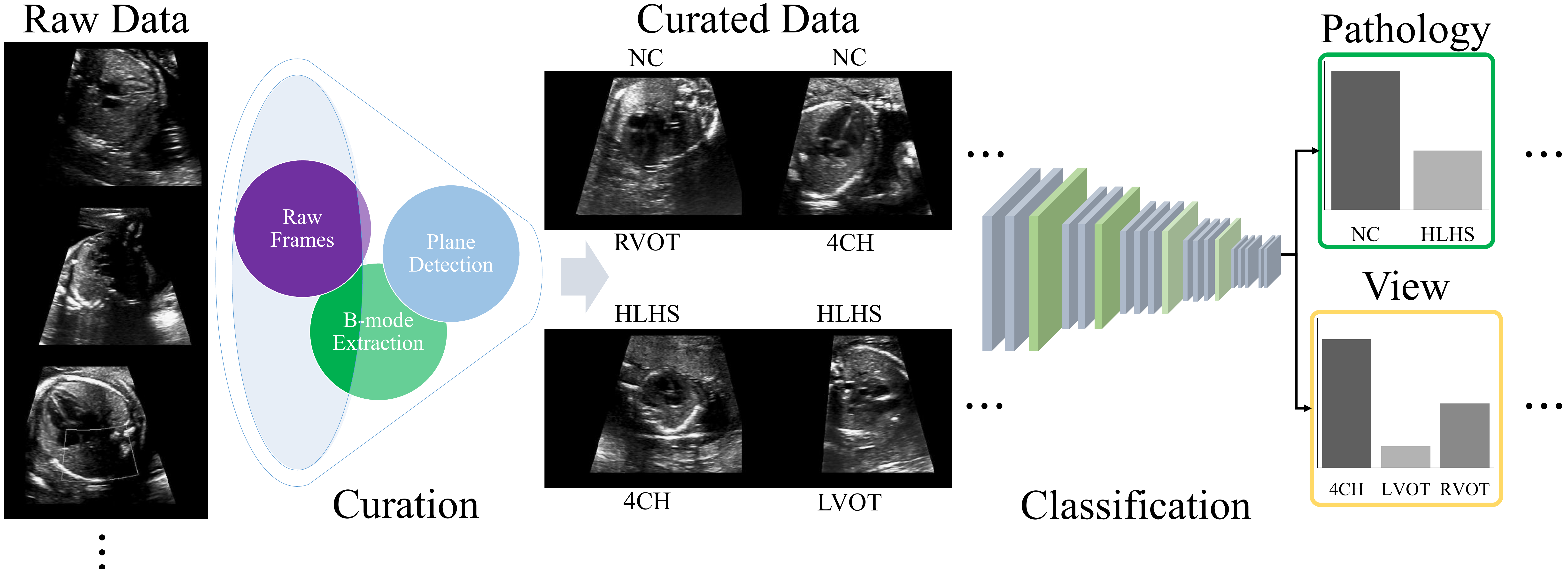}
		\caption{\textit{Automated curation and classification pipeline.}}
	\label{figures:network}
\end{figure*}

The proposed pipeline consists of two key stages, plane extraction and pathology classification. Given uncurated data, this automated workflow allows for a model to be trained for CHD classification. Figure~\ref{figures:network} depicts the overall workflow. 

\subsection{Data Characteristics}
Data was collected from 39 healthy patients and 61 patients diagnosed with HLHS, for a total of 100 patients. Each patient's data is collected as raw ultrasound DICOM videos. These videos are unannotated, meaning they contain i) frames of irrelevant anatomy, ii) frames in the vicinity of the heart, and iii) precise standard planes which are typically used for diagnosis. Table~\ref{table:data_profile1} summarizes the distribution of frames. The final extracted frames used for training are greyscale images of dimensions 224x188 pixels. The curation process used to go from raw videos to extracted frames is described in the following section.

\begin{table*}[!htbp]
    \caption{Distribution of patients, clean cardiac frames, and different cardiac views.}
	\label{table:data_profile1}
	\renewcommand{\arraystretch}{1.3}
	\centering
	\resizebox{\textwidth}{!}{
	\begin{tabular}{lrrrrrr|}
		\cline{2-7}
		\multicolumn{1}{l|}{} &  \multicolumn{3}{c}{\textbf{NC}} & \multicolumn{3}{c|}{\textbf{HLHS}} \\ \cline{2-7} 
		\multicolumn{3}{c}{} \\[-1\normalbaselineskip] \hline
		\multicolumn{1}{|l|}{Unique Patients}	&  \multicolumn{3}{c|}{39} & \multicolumn{3}{c|}{61} \\ 
		\multicolumn{1}{|l|}{Total frames in DICOM files}	& \multicolumn{3}{c|}{354867} & \multicolumn{3}{c|}{741290} \\
		\multicolumn{1}{|l|}{Clean frames after curation}	&  \multicolumn{3}{c|}{189000} & \multicolumn{3}{c|}{376337} \\ 
		\multicolumn{1}{|l|}{Clean frames (cardiac views)}	&  \multicolumn{3}{c|}{102993} & \multicolumn{3}{c|}{143468} \\ \hline
		\multicolumn{1}{|l|}{View}	&  \multicolumn{1}{c}{4CH} & \multicolumn{1}{c}{LVOT} & \multicolumn{1}{c|}{RVOT} &
		\multicolumn{1}{c}{4CH} & \multicolumn{1}{c}{LVOT} & \multicolumn{1}{c|}{RVOT} \\ \hline
		\multicolumn{1}{|l|}{Extracted Frames}	&  
		31938 & 62195 & \multicolumn{1}{c|}{8452} &
		49176 & 73203 & \multicolumn{1}{c|}{19959} \\
		
		\hline
	\end{tabular}
	}
	
\end{table*}

\subsection{Plane Extraction}
The data curation pipeline involves i) extraction of B-mode frames, and ii) standard plane extraction. In the B-mode extraction phase, frames are run through a series of tests which detect the presence of certain colors, user interface elements, or particular histogram characteristics. These help remove Doppler, split-view, and M-mode frames respectively. 

A standard plane detector, SonoNet~\cite{baumgartner2017sononet}, is used to extract relevant cardiac frames. Specifically, the 4 chamber heart (4CH), left ventricular outflow tract (LVOT) and right ventricular outflow tract (RVOT) views are used. Table~\ref{table:data_profile1} displays the number of frames extracted for each view. Note that this standard plane detector has been trained to detect cardiac frames in \textit{healthy} patients. As such, its ability to detect relevant frames in patients with malformed hearts may be impaired. Given the bias toward normal hearts, the detector may not recognize frames which show gross defects, which would be the most diagnostically relevant. Instead it may favor instances where the heart appears closer to normal, potentially making them more difficult to distinguish. While this is unideal, it represents the most general case where we do not have access to annotations which highlight the most diagnostic frames. This circumvents the need to train individual plane detectors for every pathology. To test whether diagnostic information can be gleaned from these images, we train a classifier as described in the following section.


\subsection{CHD Classification}
We train a single classifier to discriminate between healthy and HLHS patients. All three cardiac views are used for training within the same network. The network architecture is the same as SonoNet~\cite{baumgartner2017sononet} which has been inspired by the staple VGG network~\cite{simonyan2014very}. A standard binary cross-entropy loss (Eq.~\ref{equation:loss_CHD}) is used for optimization. 

In the low data regime (particularly when the number of unique patients is low) it is difficult to ensure that the classifier learns features that are genuinely related to pathology. Instead, the network might learn extraneous features that are reliable within the training data but are not robust within the test set. It is also prohibitively expensive to annotate which regions are important in each image. As such, we exploit the view labels which come for free from the data curation pipeline. Discriminating between the 4CH, LVOT and RVOT views requires the network to identify cardiac structures. It is also an intra-patient task, meaning that patient-specific features are not reliable. View classification is thus added as auxiliary task to bias the network toward cardiac structures and away from memorization of patient-specific characteristics that do not generalize to test data. The auxiliary loss and combined multitask~\cite{caruana1997multitask} loss are given in Eq.~\ref{equation:loss_view} and Eq.~\ref{equation:loss_multitask} respectively. 

\begin{equation}
\mathcal{L}_{\textrm{CHD}}(x_i,y_i,f) = -\sum_{c=1}^{N=2}y_{i,c}\textrm{log}(f(x_{i,c}))
\label{equation:loss_CHD}
\end{equation}

\begin{equation}
\mathcal{L}_{\textrm{view}}(x_i,v_i,f) = -\sum_{c=1}^{N=3}v_{i,c}\textrm{log}(f(x_{i,c}))
\label{equation:loss_view}
\end{equation}

\begin{equation}
\mathcal{L}_{\textrm{Multitask}}(x_i,y_i,v_i,f) = \mathcal{L}_{\textrm{CHD}} + \lambda\mathcal{L}_{\textrm{view}}
\label{equation:loss_multitask}
\end{equation}

\begin{equation}
\lambda_i = 1 -0.5\mathcal{L'}_{\textrm{CHD}}(x_i,y_i,f)
\label{equation:multi-taskWeighting}
\end{equation}

The weight of the view loss, $\lambda$, is computed individually for each instance (Eq.~\ref{equation:multi-taskWeighting}). It is based on the minmax-scaled CHD loss values, $\mathcal{L'}_{\textrm{CHD}} \in [0,1]$, of each sample within a batch. The view loss only increases when a sample has a lower CHD loss, making CHD the network's priority. There are more sophisticated ways of finding the optimal weighting between tasks, e.g.~\cite{kendall2018multi}, but this simplistic approach has little overhead and demonstrates an improvement over a naive setting of $\lambda=1$.    


\subsection{Assessing Diagnostic Quality during Inference}
\label{section:robustInference}
The automated curation pipeline helps to reduce manual data processing. However, it may extract low quality frames which have less diagnostic information. Including predictions on such frames adds noise to the overall diagnosis. As such, we aim to identify and remove unreliable predictions. Note that even if \textit{all} frames from a patient are rejected, it is better than providing a diagnosis which is unreliable. 


A reliable prediction for cardiac disease should depend on the cardiac structures themselves, rather than extraneous features in the background. Such a prediction should be robust to small perturbations as long as the cardiac structures are kept intact. Our aim is to generate these perturbations and use them to identify which predictions are reliable. 

Generating perturbations that preserve the cardiac structures requires knowledge of the location of the cardiac structures. However, annotating these structures is expensive. As such, we exploit the auxiliary view task as a proxy. The view classification task determines which cardiac plane is presented in the image, which depends heavily on the cardiac structures. To generate a perturbation we use the gradient from the view task to distort the image in a way that does not change the the view prediction (i.e. the cardiac structures). This is similar to adversarial examples~\cite{Szegedy2014Intriguing}, except we keep the prediction the same (Eq.~\ref{equation:adv_perturbation}) by following the negative gradient (Eq.~\ref{equation:adv_step}~\cite{madry2018towards}). Multiple perturbation steps, $\delta$, can be taken; each time using a step size of $\alpha=2$ up to a maximum perturbation of $\pm8$ from a pixel's original value $p\in[0,300]$.



\begin{equation}
\min_{\delta\in\Delta}
\mathcal{L}_{\textrm{view}}(x_i+\delta,v^*_i,f) \;,\; 
\textrm{where} \; v^*_i=f(x_i)
\label{equation:adv_perturbation}
\end{equation}
\begin{equation}
\delta = -\alpha \textrm{sgn}(\nabla_x\mathcal{L}_{\textrm{view}}(x_i,v^*_i,f))
\label{equation:adv_step}
\end{equation}

\subsection{Evaluation}
After automatically extracting the relevant frames for all patients, \textit{manual} curation was performed on images from roughly 20\% of the patients. In manual curation, the images are sorted into three groups based on their diagnostic quality (high, medium, and low) by expert clinicians. Half of these patients are used for testing and the other half are used for validation. Evaluation on the test data measures how well the model is able to learn generalizable features from the automatically curated data. The quality of the automated curation can also be measured by comparing results on test sets with and without manual curation. We report precision, recall, F1-scores and the area under curve for the receiver operator characteristic curve (ROC-AUC).

\section{Results}

An overview of the results is presented in Table~\ref{table:Ablation}. The test data produced by automated curation contains images with varying levels of diagnostic quality. Testing on all levels (low-high quality) results in poor performance. Using medium-high quality immediately improves all metrics. The multitask loss provides a considerable improvement and the weighted multitask approach ($\lambda \propto \mathcal{L}_{\textrm{CHD}}$, Eq.~\ref{equation:multi-taskWeighting}) further improves performance. Using robust inference (described in Section~\ref{section:robustInference}) helps to remove predictions that are deemed less reliable. The fact that performance increases indicates that robust inference is effective in identifying and excluding predictions that are less accurate.


\begin{table*}[!hbtp]
    \caption {Pathology classification results for different models using testing data with different diagnostic quality.}
	\label{table:Ablation}
	
	\renewcommand{\arraystretch}{1.2}
	\centering
	\resizebox{1.\textwidth}{!}{
	    	\begin{tabular}{llllllllllll}
			\cline{1-12}
			\multicolumn{1}{|l}{\textbf{Loss}} & 
			\multicolumn{1}{p{0.1\linewidth}}{\textbf{Test Quality}} & 
			\multicolumn{2}{p{0.14\linewidth}}{\textbf{Precision (NC:CHD)}} & 
			\multicolumn{2}{p{0.14\linewidth}}{\textbf{Recall (NC:CHD)}} & 
			\multicolumn{2}{p{0.14\linewidth}}{\textbf{F1-score (NC:CHD)}} & 
			\multicolumn{1}{p{0.08\linewidth}|}{\textbf{ROC-AUC}} \\ 
			\cline{1-12}
			
			\multicolumn{6}{c}{} \\[-1\normalbaselineskip] \hline
			
			\multicolumn{1}{|l}{CHD} & All &  0.72 & 0.64 & 0.59 & 0.76 & 0.65 & 0.70 & \multicolumn{1}{l|}{0.75} \\
			
			\multicolumn{1}{|l}{CHD} & Med-High & 0.72 & 0.77 & 0.71 & 0.77 & 0.72 & 0.77 & \multicolumn{1}{l|}{0.82} \\ 
			\multicolumn{1}{|l}{Multitask ($\lambda=1$)} & Med-High & 0.77 & 0.81 & 0.77 & 0.81 & 0.77 & 0.81 & \multicolumn{1}{l|}{0.87} \\ 
			\multicolumn{1}{|l}{Multitask ($\lambda \propto \mathcal{L}_{\textrm{CHD}}$)} & Med-High & 0.80 & 0.80 & 0.74 & 0.85 & 0.77 & 0.83 & \multicolumn{1}{l|}{0.89} \\
			\multicolumn{1}{|l}{Multitask ($\lambda \propto \mathcal{L}_{\textrm{CHD}}$)} & Robust Frames & 0.83 & 0.89 & 0.90 & 0.82 & 0.87 & 0.85 & \multicolumn{1}{l|}{0.93} \\
			\hline
		\end{tabular}
	}
	
\end{table*}




\begin{figure*}[h]
	\centering
        \includegraphics[width=0.8\linewidth]{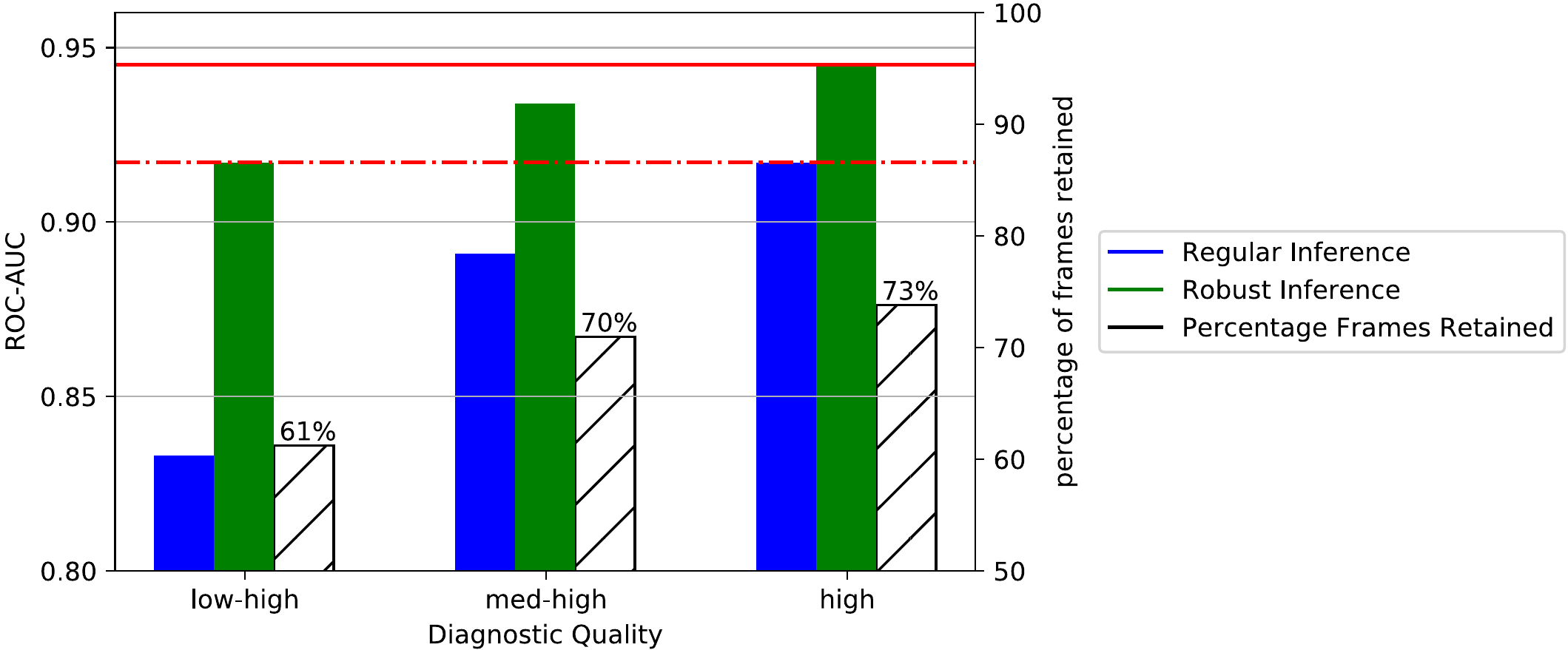}
		\caption{\textit{ROC-AUC comparison of regular and robust inference. Including low quality frames quickly leads to a sharp decline in performance (blue). Robust inference can help to identify frames that are more reliable (white hatch, right y-axis). Evaluating only reliable frames leads to more accurate overall predictions (green).}}
	\label{figures:robustInference}
\end{figure*}
A closer examination of robust inference is given in Figure~\ref{figures:robustInference}. Using regular inference (blue), the ROC-AUC becomes severely compromised when including low quality images. However, the performance can be recovered by using robust inference (green) to automatically determine when a prediction is less reliable. 

The white hatched bars (Figure~\ref{figures:robustInference}) indicate the percentage of images which are deemed reliable in each set. The percentage of reliable images increases as the expert-rated quality increases. This indicates that the proposed reliability rating corresponds to expert judgement to some extent. Ideally 100\% of the high quality images should be retained; however 73\% are deemed reliable. Nonetheless the excluded predictions were indeed unreliable and their exclusion results in an improvement of almost 0.03  (solid and dotted red lines). In the low-high set, only 61\% of frames are retained. To put this in perspective, experts only considered 39\% of the low-high set as being of acceptable quality (medium or above).

\begin{figure*}[h]
	\centering
        \includegraphics[width=0.5\linewidth]{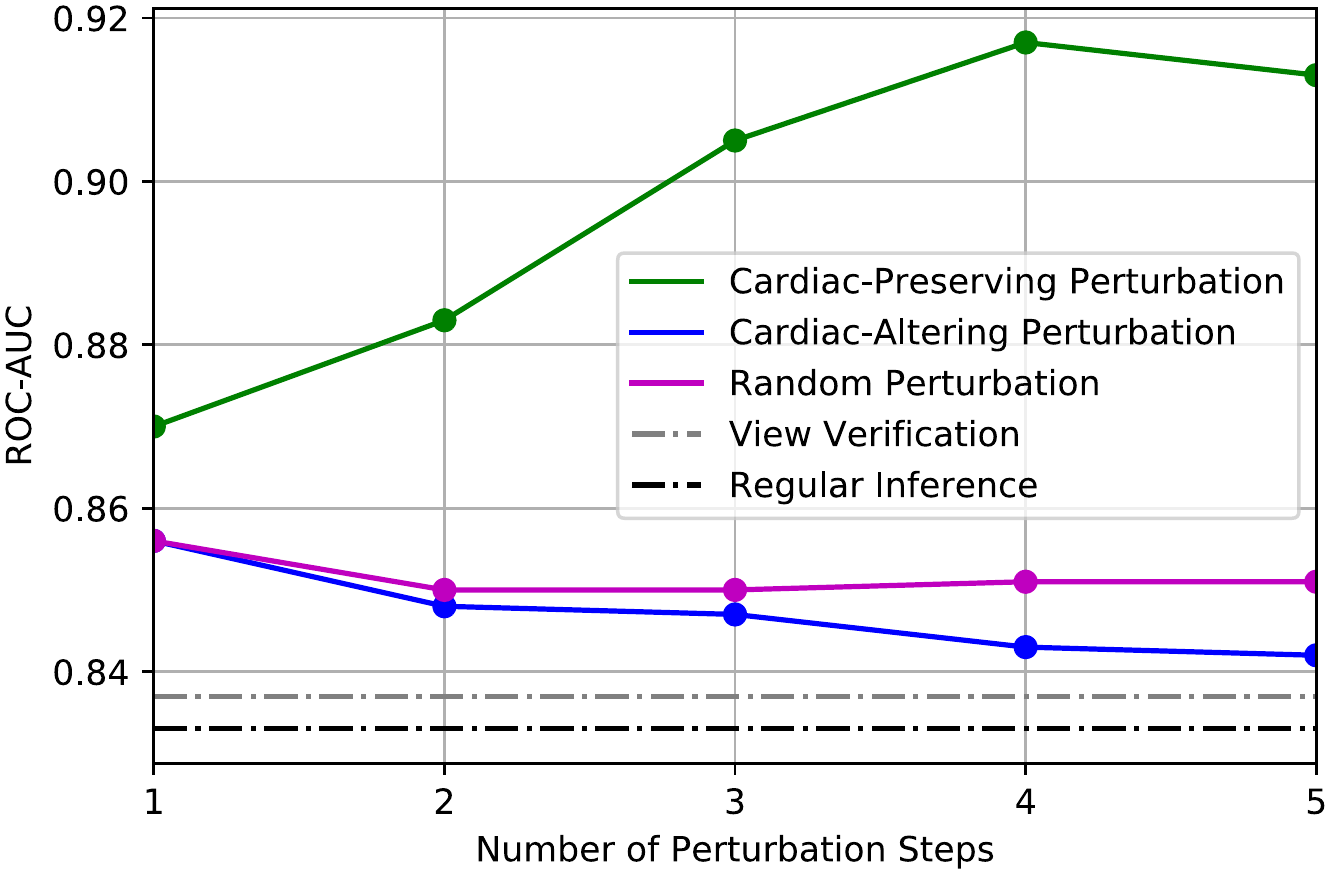}
		\caption{\textit{ROC-AUC using different approaches of frame quality assessment. All methods improve upon regular inference (black). Cardiac-preserving perturbations stand out as being considerably better at removing unreliable predictions (green).}}
	\label{figures:perturbations}
\end{figure*}

We also compare different approaches to robust inference in Figure~\ref{figures:perturbations}. The view verification approach excludes any images which produce an incorrect view prediction (based on `ground truth' view labels provided by the plane detector in the curation pipeline). This gives very limited improvement (grey). 

The proposed cardiac-preserving perturbations are able to find and remove inaccurate predictions, which consistently improves ROC-AUC (Figure~\ref{figures:perturbations}, green). In fact, increasing the number of perturbation steps actually leads to better results. 
In comparison, random (purple) and cardiac-altering perturbations (i.e. adversarial examples - blue) do not provide as much improvement. 






\section{Discussion}
Various elements, including multitask learning~\cite{caruana1997multitask} and robust inference, contribute toward improving the ROC-AUC from 0.75 to over 0.91. With a small number of unique patients and a lack of manual curation, the standard classifier performs poorly. Multitask learning~\cite{caruana1997multitask} helps to provide a bias toward cardiac structures that are relevant for the view classification task. These features are also helpful for pathology classification and lead to an increase in performance (Table~\ref{table:Ablation}). 

Robust inference also helps to improve the scores by filtering out unreliable predictions. Predictions are considered less reliable if they change when the image is perturbed in a way that preserves cardiac features. Stronger perturbations (using more steps) are more likely to alter predictions. However, predictions that rely on cardiac features should remain unaffected. Figure~\ref{figures:perturbations} demonstrates that strong cardiac-preserving perturbations are the most adept at finding unreliable predictions. In comparison, random and cardiac-altering (adversarial) perturbations are not able to tease apart reliable and unreliable predictions. 

We find that robust inference is an important component because the automated curation pipeline includes many images of low diagnostic quality. Typically, these frames would have to be manually vetted in order to reduce the noise in the overall prediction. However, robust inference can serve as a filter to remove predictions from low quality images. For low-high quality images, robust inference improves the ROC-AUC from 0.83 to 0.92. This is on par with the ROC-AUC achieved with regular inference on \textit{only} high quality images (0.92). This is highlighted in Figure~\ref{figures:robustInference} with a red dotted line.  

Measuring the reliability of a prediction is similar to uncertainty estimation (e.g.~\cite{gal2016dropout}). However, uncertainty quantification is a much more complex task and is not always straightforward to interpret~\cite{yao2019quality}. Instead, we simply measure the prediction's dependence on cardiac structures (which should hold key diagnostic information). With this prior knowledge we can estimate \textit{our} confidence in the prediction, rather than the network's intrinsic uncertainty.  
 
We demonstrate that it is possible to train a classifier to identify CHD from standard B-mode images. In the future, temporal information or Doppler images could be used to provide more diagnostic information. Nevertheless, this work represents a step toward the goal of assisting front-line sonographers with CHD diagnosis at a population level.


\section{Conclusion}
We propose an automated pipeline for curating and classifying CHD in fetal ultrasound. The curation step extracts relevant frames which makes training possible. Automated curation includes many images with poor diagnostic quality. Noisy predictions on such images can lead to a reduction in overall performance. As such, robust inference is introduced in this work to exclude predictions which may be less trustworthy. This helps to achieve a F1-score of 0.87 and 0.85 for NC and CHD classes respectively and an ROC-AUC of 0.93. 

Acknowledgements: Support from Wellcome Trust IEH Award iFind project [102431]. JT is supported by the ICL President's Scholarship. 

%
%
%
\bibliographystyle{splncs04}


\end{document}